\documentclass[prc,superscriptaddress,showpacs,amssymb,amsmath,
amsfonts,aps,twocolumn,secnumarabic,floatfix]{revtex4}
\usepackage{graphics}
\usepackage{epsfig}

\begin{document}
\bibliographystyle{try} 
 
\topmargin 0.1cm 
 
\title{Rescattering in Meson Photo-production off Few Body Systems}

\newcommand*{\JLAB }{ Thomas Jefferson National Accelerator Facility, Newport News, Virginia 23606} 
\affiliation{\JLAB } 

\author{ J.-M. Laget }
     \affiliation{\JLAB}

\date{\today}

\begin{abstract}
Exclusive reactions induced at high momentum transfer in few body
systems provide us with an original way to study the production and propagation of hadrons in cold nuclear matter. In very well defined parts of the phase space, the reaction amplitude develops a logarithmic singularity. It is on solid ground since it depends only on on-shell elementary amplitudes and on low momentum components of the nuclear wave function. This is the best window to study  the propagation of exotic configurations of hadrons such as, for instance, the onset of color transparency. It may appear earlier in meson photo-production reactions, more particularly in the strange sector, than in more classical quasi elastic scattering of electrons. More generally, those reactions provide us with the best tool to determine the cross section of the scattering of various hadrons (strange particles, vector mesons) from the nucleon and to access the production of possible exotic states. 
\end{abstract}


\maketitle

\section{Introduction}
Exclusive reactions induced at high momentum transfer in few body
systems provide us with an original way to study the production and propagation of hadrons in cold nuclear matter. In very well defined parts of the phase space, the reaction amplitude develops a logarithmic singularity which enhances the cross section. Here, the reaction amplitude is on solid ground since it depends only on on-shell elementary amplitudes and on low momentum components of the nuclear wave function~\cite{La81}. On the one hand this allows the determination of the scattering cross section of short lived particles. On the other hand this allows the study of the propagation of exotic configurations such as the onset of Color Transparency. 

The  concept  of  Color  Transparency  follows  from  the underlying
structure  of  QCD:    interactions  between ``white'' objects depend on
their  transverse  size~\cite{MuXX,Fa90}.    A  hard  scattering  of  the probe produces recoiling particles  with small  transverse size,  whose the  subsequent interactions in  Nuclear Matter  are reduced.   There  is no  doubt that Color Transparency should occur.  The question is where and when.

The difficulty resides in the fact that such an exotic configuration evolves quickly toward the asymptotic state of the detected hadron: in order to observe Color Transparency, the characteristic scale of this evolution should be larger,  or comparable, to the size of the largest nuclei.

To date there is no  convincing evidence for Color Transparency in photon and electron induced reactions.   The
reason is  that most  of the  attempts were  performed in semi-exclusive
kinematics.  In the A(e,e'p) reactions~\cite{EnXX,On95} for instance, it is very likely that the values of the only available hard scale, $Q^2$, are too low  to observe color transparency  in
the  quasi-free  kinematics  channels,  where  the energy of the ejected
nucleon  $T_p$  and  the   photon  four-momentum  are  not   independent
($T_p=Q^2/2M$).  In the range  of values of $Q^2$ accessible today, the life time of the small object  is
of the order  of the distance  between nucleons rather  than the nuclear
radius.  For instance, at the highest $Q^2= 6$ GeV$^2$ where data exist,
the energy of the outgoing proton  is only 3 GeV and its  characteristic
evolution distance~\footnote{An educated guess of the formation length is $l_f=2\,\hbar c\,p/\Delta m^2$, with $\Delta m^2\sim 0.8$~GeV$^2$. When $p\sim T_p=3$~GeV, one gets $l_f=1.5$~fm} is no more than 1.5 fm, closer to the internucleonic distance rather than the size of the nucleus.

A signal has been reported in A(e,e'$\rho$) reactions, at Fermi Lab.~\cite{FNAL} and DESY~\cite{HERMES}. However,  it comes from a subtle interplay between the attenuation of the hadronic component of the virtual photon in the entrance channel and the onset of color transparency in the exit channel (see e.g.~\cite{Ko02}). An experiment~\cite{HaXX} has been completed recently at Jefferson Laboratory (JLab) to disentangle these two effects: one has to wait the final analysis for a more definite answer.

The way  to overcome  these difficulties  is to  study reactions induced by
photons in few body systems:  exclusive reactions allow the  formation  length  of  the  hadron to be adjusted to the distance between
nucleons~\cite{La98,Fr97,Fr96}.  The kinematics should be chosen so that the interactions of the emerging hadron with a second nucleon are maximal. This occurs when the  produced hadron  propagates   on-shell and rescatters on  a second nucleon at rest (triangular   logarithmic singularity).    A  clear  signal  for  color  transparency would be the suppression  of the  final  state  interaction peak  when  the momentum transfer increases. This situation is more comfortable  than in the more classical  study of quasi elastic scattering of electrons from heavy nuclei, where one  looks for a change  of a flat  level of attenuation  of the outgoing  nucleon, instead of the evolution of a well defined peak.

This conjecture~\cite{La98,Fr97,Fr96} has been tested in two studies of the $^2$H$(e,e^{\prime}p)n$ reaction that have been completed recently at JLab: the first~\cite{BoXX} with two magnetic spectrometers in Hall A; the second~\cite{EgXX} with the CEBAF Large Acceptance Spectrometer~\cite{CLAS} (CLAS) in Hall B. The preliminary results do not exhibit a signal of color transparency (within the experimental and theoretical uncertainties) in the range $2<Q^2<6$~GeV$^2$, when compared to the latest prediction~\cite{La05} of the various interaction mechanisms.

It may  occur earlier  in Exclusive Photo-production  of Mesons.   The
reason is that mesons are made up of two quarks. They may recombine more easily, through the exchange of one hard gluon only, and prepare a configuration with a small transverse extension which further evolves toward its asymptotic size. The second reason is that the hard scale is provided by the four momentum transfer $t$ which happens to define the interaction volume, while the virtuality $Q^2$ of the photon that is exchanged in $(e,e'p)$ reactions defines the scale of observation~\cite{La04}.

Indeed, a hint has been reported in the $^4$He($\gamma$,p$\pi^-$) and $^{16}$O($\gamma$,p$\pi^-$) channels recently studied at JLab~\cite{Du03,Ga96}. However, the signal is weak and again these are semi exclusive reactions. The signal must be confirmed by completely exclusive measurements.

\begin{figure}[h]
\begin{center}
\epsfig{file=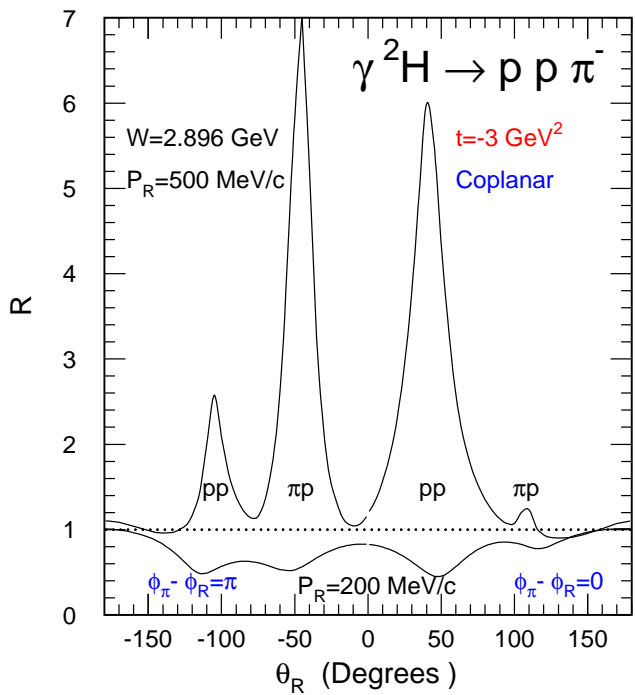,width=3.5in}
\caption{Color on line. Ratio of the total
to the quasi-free cross section  of the $^2$H$(\gamma,\pi^-p)p$ reaction
against the angle of the recoiling proton whose the momentum is kept
constant at 500 MeV/c (top) and 200 MeV/c (bottom). The peaks labeled $\pi p$ and $pp$  correspond respectively  to $\pi p$ and $pp$ on shell rescattering. The dotted curve corresponds to the quasi-free process. The kinematics is coplanar, and positive angles correspond to the emission of the pion and the recoiling proton on the same side of the photon.}
\protect\label{pi1}
\end{center}
\end{figure}

The simplest example is the  reaction $^2$H$(\gamma,p\pi^-)p$ in the energy  range
$4 < E_{\gamma} < 10$ GeV. For real photons the momentum  transfer $t$,
between the incoming photon and the outgoing pion, sets the size of  the
interaction volume.   As  can be  seen in  Fig.~\ref{pi1}, the  on shell
rescattering peaks  corresponding to  $\pi p$  or $pp$  interactions are
clearly separated. Such  a  Logarithmic  singularity  has  already  been  observed at lower
energies~\cite{Ar78}. At the top of each peak, the rescattering  amplitude
is dominated by  low momentum components  of the deuteron  wave function
and on mass shell elementary reaction amplitudes (see Ref.~\cite{La81}).
The elementary reaction
$n(\gamma,\pi^-)p$ is  well  reproduced  by  a  model based on the
exchange of saturating Regge trajectories~\cite{La97}. The $\pi$-nucleon, as well as the nucleon-nucleon, elementary scattering amplitudes are almost entirely absorptive, and well under control at high energy~\cite{La72,PDG}. The rescattering amplitudes are therefore on solid grounds~\cite{La05}: the method can be calibrated at low values of the four momentum transfer $t$. A signal of color transparency would be the reduction of the $\pi p$ rescattering peak when $t$ increases. It could happen sooner in the strange sector, where the strange quark may play a special role. 

Alternatively, the method will allow a determination of the cross section of hyperons ($\Lambda$, \ldots) or vector mesons ($\phi$, J/$\Psi$, \ldots) scattering with nucleons, or the interactions between other unstable hadrons. More generally it offers us a way to access the mechanisms of the formation of hadrons in cold nuclear matter.

I have already presented these conjectures in several conference talks~\cite{La98,La98a,La00} and prospect reports.
In the mean time the CLAS  collaboration at JLab has completed, with an unprecedented statistics, a study~\cite{g10} of the interactions of a real photon beam (maximum energy 3.7 GeV) on a deuterium target: it provides us with a unique testing ground of these ideas and the method.

This article is not one more attempt to predict the size of the signal of Color Transparency: too many parameters are unknown and although we already know what are the qualitative expectations, only experiments will allow us to quantify the effect. This article is rather an attempt to provide a comprehensive recollection and update of the various matrix elements in the meson production sectors and to provide a base line calculation in a dedicated kinematical range, already accessible at JLab at 6 GeV and its further upgrade to 12 GeV, from which any deviation will be meaningful. The nucleon sector has already been revisited in ref.~\cite{La05}. Section~\ref{pi} deals with the pion production sector, section~\ref{strange} deals with the single kaon production sector, section~\ref{vector} deals with the vector meson production sector (more  specifically $\phi$ and J/$\Psi$), section~\ref{ct} addresses issues in Color Transparency, section~\ref{pent} investigates possible implication in the search of exotics, while section~\ref{conc} concludes and summarizes the prospects.

\section{The $^2$H$(\gamma,\pi^-p)p$ reaction \label{pi}}

\subsection{The model \label{model}}

The model is a straightforward update of the diagrammatic approach~\cite{La81} which has been successful in the analysis of meson production reactions at lower energies (let's say in the resonance region). It is particularly well suited to evaluate the reaction amplitude near the singularities of the $S$-matrix. The kinematics, the elementary operators as well as the propagators are relativistic. The deuteron wave function corresponds to the Paris potential~\cite{PaXX}, but any modern wave function leads to very similar results in the momentum range covered by this study. 

Let $k=(\nu,\vec{k})$, $p_D=(M_D,\vec{0})$, $p_{\pi}=(E_{\pi},\vec{p_{\pi}})$, $p_1=(E_1,\vec{p_1})$ and $p_2=(E_2,\vec{p_2})$ be the four momenta, in the Lab. system, of respectively the incoming photon, the target deuteron, the outgoing pion, the slow outgoing proton and the fast outgoing proton. The 5-fold fully differential cross section is related to the square of the coherent sum of the matrix elements as follows:
\begin{eqnarray}
\frac{d\sigma}{d\vec{p_1}[d\Omega_{\pi}]_{cm2}}&=& \frac{1}{(2\pi)^5}
\frac{|\vec{\mu}_{c.m.}|m^2}{24|\vec{k}|E_1Q_f}
\sum_{\epsilon,M,m_1,m_2}\left| \sum_{i=I}^{III}
\right.\nonumber \\
&&{\cal M}_i(\vec{k},\epsilon,M,\vec{p_{\pi}},\vec{p_1},m_1,\vec{p_2},m_2)
\nonumber \\
&&\left.
-{\cal M}_i(\vec{k},\epsilon,M,\vec{p_{\pi}},\vec{p_2},m_2,\vec{p_1},m_1)
\right|^2
\label{cross}
\end{eqnarray}
where $\epsilon$ is the polarization vector of the photon and where $M$, $m_1$ and $m_2$ are the magnetic quantum numbers of the target deuteron and the two outgoing protons respectively. The norm of the spinors is $\overline{u}u=1$. The amplitudes are computed in the Lab. frame. The antisymmetry between the two outgoing protons is insured by exchanging the role of $(\vec{p_1}, m_1)$ and  $(\vec{p_2}, m_2)$ in the second amplitude. The cross section is differential in the Lab. three-momentum of proton 1, but in the solid angle of the pion expressed in the c.m. frame of the pair made by the pion and proton 2. In this frame, the momentum of the pion is $\vec{\mu}_{c.m.}$ and the total energy is $Q_f= \sqrt{(E_{\pi}+E_2)^2 -(\vec{p_{\pi}} +\vec{p_2})^2}$.

\begin{figure}[h]
\begin{center}
\hspace{-2cm}
\epsfig{file=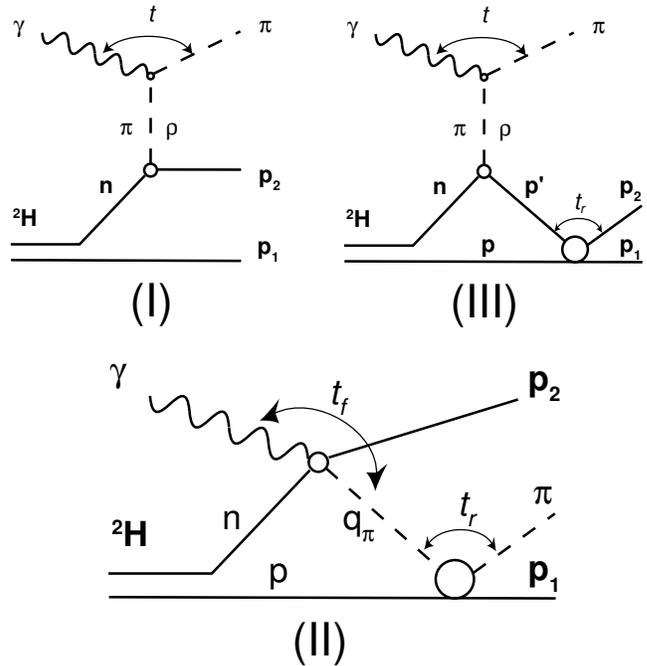,width=4.in}
\caption{
The relevant mechanisms. I: Quasi-free. II: Meson-Nucleon rescattering. III: Nucleon-Nucleon rescattering}
\protect\label{graph}
\end{center}
\end{figure}

The cross section and the amplitudes are given for the case of a real photon induced reaction, which I consider in this study. They depend only on the transverse components $J_X$ and $J_Y$ of the hadronic current $\vec{J}$. In the case of a virtual photon beam, additional terms in the cross section are related, as outlined in ref.~\cite{La94}, to the longitudinal component of the hadronic current, $J_z$.  The matrix elements  are expressed as the scalar product $\vec{J}\cdot\vec{\epsilon}$ from which each component of the hadronic current can be deduced. In the following, I give the expressions of the dominant reaction amplitudes in Fig.~\ref{graph} and discuss their update relevant to the high energy domain accessible at JLab. I refer the reader to~\cite{La78} for more technical details.

\subsubsection{Quasi-free meson production}

\begin{figure}[h]
\begin{center}
\epsfig{file=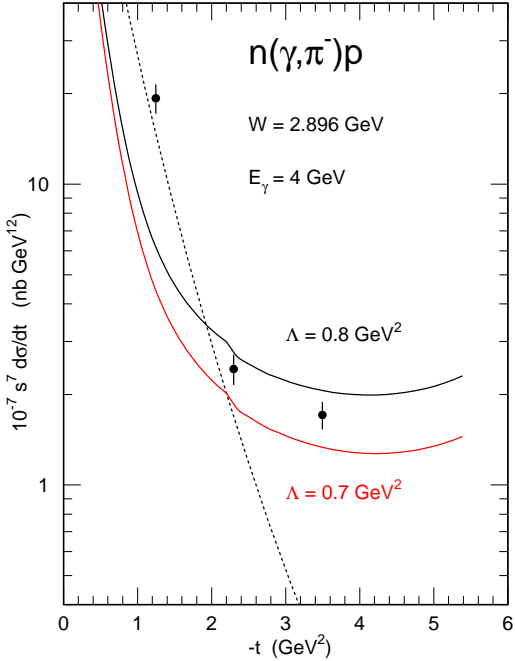,width=3.in}
\caption{Color on line.
The cross section of the elementary reaction $n(\gamma,\pi^-)p$ at $E_{\gamma}= 4$~GeV. The curve is the prediction of the Regge model. The dashed curve corresponds to linear trajectories. The full curves correspond to saturating trajectories, for two choices of the cut-off mass in the hadronic form factor. The data has been recently recorded at JLab~\cite{Zh04}.}
\protect\label{pi-}
\end{center}
\end{figure}

The matrix element of the quasi-free amplitude (graph I in Fig.~\ref{graph}) takes the simple form:
\begin{eqnarray}
{\cal M}_I(\vec{k},\epsilon,M,\vec{p_{\pi}},\vec{p_1},m_1,\vec{p_2},m_2)=
\nonumber \\
i \sum_{m_n m_l m_s} \sum_{ls} (lm_lsm_s|1M)(\frac{1}{2}m_n\frac{1}{2}m_1|sm_s)
\nonumber \\
u_l(|\vec{p_1}|) Y_l^{m_l}(\vec{p_1})
T_{\gamma n}(\vec{p_2},m_2,-\vec{p_1},m_n)
\label{q_f}
\end{eqnarray}
where $u_0$ and $u_2$ are the $S$ and $D$ components of the deuteron Paris wave function~\cite{PaXX}, and where $T_{\gamma n}$ is the amplitude of the elementary $n(\gamma,\pi^-)p$ reaction.  I use the on-shell expression (see Appendix) of the Regge amplitude of ref.~\cite{La97}, which is based on the exchange of the saturating Regge trajectories of the pion and the rho mesons. It leads to a good description of the differential cross section of the $p(\gamma,\pi^+)n$ reaction at large momentum transfer $-t$ in the photon energy range of JLab (around 4~GeV). As shown in Fig.~\ref{pi-}, it leads also to a fair accounting of the more recent JLab data~\cite{Zh04} in the  $\pi^-$ channel. I refer the reader to~\cite{La97} for a throughout presentation of this Regge model and the choice of the coupling constants and parameters: I use the same in this study, except for the cut-off mass of the hadronic form factor, which I chose to be $\Lambda=0.7$~GeV$^2$, instead of 0.8~GeV$^2$ in ref.~\cite{La97}.  

When the momentum, $\vec{p_1}$, of one of the proton is low only one amplitude dominates the cross section~(\ref{cross}), which takes the simple form~\cite{La81,La78}:
 \begin{eqnarray}
\frac{d\sigma}{d\vec{p_1}d\Omega_{\pi}}=
(1+\beta_1 \cos\theta_1)\rho(|\vec{p_1}|)\frac{d\sigma}{d\Omega_{\pi}}
(\gamma n\rightarrow \pi^-p)
\label{cross_qf}
\end{eqnarray}
where $\beta_1=p_1/E_1$ and $\theta_1$ are the velocity and the angle of the spectator nucleon. This is nothing but the relation between the yield and the elementary cross section of the production of a pion on a nucleon which moves with the velocity $-\vec{\beta_1}$. The number of target nucleons is $\rho(|\vec{p_1}|)d\vec{p_1}$, where $\rho(|\vec{p_1}|)$ is the momentum distribution of the neutron in deuterium, while $(1+\beta_1 \cos\theta_1)$ is the flux of photons seen by the moving target nucleon.

\subsubsection{Meson-nucleon rescattering}

The matrix element of the pion-proton rescattering amplitude (graph II in Fig.~\ref{graph}) takes the form:
\begin{eqnarray}
{\cal M}_{II}(\vec{k},\epsilon,M,\vec{p_{\pi}},\vec{p_1},m_1,\vec{p_2},m_2)=
\nonumber \\
i \sum_{m_n m_p} (\frac{1}{2}m_n\frac{1}{2}m_p|1M)
\int \frac{d^3\vec{p}}{(2\pi)^3} \frac{u_0(p)}{\sqrt{4\pi}}
\frac{1}{q^2_{\pi}-m^2_{\pi}+i\epsilon}
\nonumber \\
\frac{m}{E_p}
T_{\gamma n}(\vec{p_2},m_2,-\vec{p},m_n)T_{\pi N}(\vec{p_1},m_1,\vec{p},m_p)
\nonumber \\
+ D \; \mathrm{wave}\ \mathrm{part} \;\;
\label{pi_rescat}
\end{eqnarray}

The integral runs on the three momentum of the spectator proton in the loop, which has been put on-shell, $p^{\circ}=E_p=\sqrt{\vec{p}^2+ m^2}$, by the integration over its energy $p^{\circ}$. It can be split in two parts:
\begin{eqnarray}
{\cal M}_{II} = {\cal M}_{II}^{on}+{\cal M}_{II}^{off}
\end{eqnarray}

The singular part of the rescattering integral runs between the minimum and maximum values of the momentum of the spectator proton in the loop for which the pion can propagate on-shell:
\begin{eqnarray}
p_{min}(p\pi)= \frac{P}{Q_s}E_{c.m.}- \frac{E}{Q_s}p_{c.m.} 
\label{pmin_ppi}
\end{eqnarray}
\begin{eqnarray}
p_{max}(p\pi)= \frac{P}{Q_s}E_{c.m.}+ \frac{E}{Q_s}p_{c.m.} 
\label{pmax_ppi}
\end{eqnarray}
where $E=E_{\pi}+E_1$, $\vec{P}=\vec{p_{\pi}}+\vec{p_1}$ and $Q_s=\sqrt{E^2-\vec{P}^2}$ are respectively the energy, the momentum and the mass of the scattering $\pi p$ pair. The momentum and energy of the spectator proton, in the c.m. frame of the $\pi p$ pair are
\begin{eqnarray}
p_{c.m.}= \frac{\sqrt{(Q^2_s-(m+m_{\pi})^2)(Q^2_s-(m-m_{\pi})^2)}}{2 Q_s}
\end{eqnarray}
\begin{eqnarray}
E_{c.m.}= \sqrt{p^2_{c.m.}+m^2} 
= \frac{Q^2_s+m^2-m^2_{\pi}}{2 Q_s}
\end{eqnarray}
 
It takes the form:
\begin{eqnarray}
{\cal M}_{II}^{on}= \frac{\pi}{(2\pi)^3\sqrt{{4\pi}}} \sum_{m_nm_p}
\frac{1}{2P} (\frac{1}{2}m_n\frac{1}{2}m_p|1M)
\nonumber \\
\int_0^{2\pi} d\phi \int_{|p_{min}(p\pi)|}^{p_{max}(p\pi)} pu_0(p) dp
\frac{m}{E_p}
\left[T_{\gamma n}T_{\pi N}\right]_{q^2_{\pi}=m^2_{\pi}}
\nonumber \\
+ D \; \mathrm{wave}\ \mathrm{part} \;\;
\label{sing_pin}
\end{eqnarray}

The two dimensional integral is done numerically. It depends only on on-shell elementary amplitudes. The weight, $pu_0(p)$, 
selects nucleons almost at rest in the deuterium when the lower bound, $p_{min}(p\pi)$, of the integral vanishes. This is the origin of the meson-nucleon scattering peak in Fig.\ref{pi1}, which is therefore on solid grounds.

The $\pi N$ scattering amplitude can be expressed as:
\begin{eqnarray}
T_{\pi N} = (m_1|f(Q_s,t_r) 
+ g(Q_s,t_r) \vec{\sigma}\cdot \vec{k}_{\perp}|m_p)
\end{eqnarray}
where $t_r=(p_{\pi}-q_{\pi})^2$ is the four momentum transfer at the $\pi p$ recattering vertex and $\vec{k}_{\perp}= \vec{p_{\pi}}\times \vec{q_{\pi}}$ is the direction perpendicular to the scattering plane. At high energies ($Q_s> 2$~GeV) the central part dominates at forward angles and is almost entirely absorptive. It can be parameterized as follows:
\begin{eqnarray}
f(Q_s,t_r)=-\frac{Q_s p_{c.m.}}{m} (\epsilon + i)\sigma_{\pi^-p} \exp[\frac{\beta_{\pi}}{2}t_r]
\label{scat_pin}
\end{eqnarray}
Above $Q_s\sim 2$~GeV, the total cross section stays constant at the value $\sigma_{\pi^-p}= 30$~mb~\cite{PDG}, and the fit of the differential cross section at forward angles leads to a slope parameter $\beta_{\pi}=6$~GeV$^{-2}$~\cite{La72}. At high energy the ratio between the real and imaginary part of the amplitude is small~\cite{PDG} and I set it to zero in this study.

With such an absorptive amplitude it is easy to see, from eqs.~(\ref{q_f})
 and~(\ref{sing_pin}), that the singular part of the rescattering amplitude interferes destructively with the quasi-free amplitude.
 
 The principal part of the rescattering integral takes the form:
\begin{eqnarray}
{\cal M}_{II}^{off}= \frac{i}{(2\pi)^3\sqrt{{4\pi}}} \sum_{m_nm_p} 
(\frac{1}{2}m_n\frac{1}{2}m_p|1M)
\nonumber \\
\oint \frac{d^3\vec{p} u_0(p)T_{\gamma n}T_{\pi N}}{q^2_{\pi}-m^2_{\pi}}
\frac{m}{E_p}
+ D \; \mathrm{wave}\ \mathrm{part} \;\;
\end{eqnarray}

It turns out that it vanishes~\cite{La78,La81} when $p_{min}(p\pi)=0$ at the top the $\pi N$ recattering peak and contributes little to its tails only, in Fig.~\ref{pi1} for instance. 

Since the Regge amplitude $T_{\gamma n}$ varies rapidly, as $s^{\alpha(t_f)}$, with the total energy $s=(E_2+E_q)^2-(\vec{p_2}+\vec{q}_{\pi})^2$ and momentum transfer $t_f=(k-q_{\pi})^2$, it can not be factorized out of the integral which should be evaluated numerically. This is not a problem for its singular part: it is a two-fold integral which involves well defined on-shell quantities. Its principal part is a three fold integral which requires a good knowledge of the off-shell extrapolation of the elementary amplitude. Since its contribution is small near the singularity, I do not take it into account in this study, in order to save time in the Monte Carlo simulation in the full phase space (section~\ref{clas}). It will be taken be into account later, in the final analysis of experimental data.

\subsubsection{Nucleon-Nucleon rescattering}

The matrix element of the proton-proton rescattering amplitude (graph III in Fig.~\ref{graph}) takes the form:
\begin{eqnarray}
{\cal M}_{III}(\vec{k},\epsilon,M,\vec{p_{\pi}},\vec{p_1},m_1,\vec{p_2},m_2)=
\nonumber \\
i \sum_{m_n m_pm'_p} (\frac{1}{2}m_n\frac{1}{2}m_p|1M)
\int \frac{d^3\vec{p}}{(2\pi)^3} \frac{u_0(p)}{\sqrt{4\pi}}
\frac{1}{{p^{\circ}}'-E'_p+i\epsilon}
\nonumber \\
\frac{m}{E_p}
T_{\gamma n}(\vec{p'},m'_p,-\vec{p},m_n)
T_{pp}(\vec{p_2},m_2,\vec{p_1},m_1,\vec{p'},m'_p,\vec{p},m_p)
\nonumber \\
+ D \; \mathrm{wave}\ \mathrm{part} \;\;
\label{p_rescat}
\end{eqnarray}

The integral runs on the three momentum of the spectator proton in the loop, which has been put on-shell, $p^{\circ}=E_p=\sqrt{\vec{p}^2+ m^2}$, by the integration over its energy $p^{\circ}$. It can be split in two parts:
\begin{eqnarray}
{\cal M}_{III} = {\cal M}_{III}^{on}+{\cal M}_{III}^{off}
\end{eqnarray}

The singular part of the rescattering integral runs between the minimum and maximum values of the momentum of the spectator proton in the loop for which the struck proton can propagate on-shell:
\begin{eqnarray}
p_{min}(pp)= \frac{P}{W}E_{c.m.}- \frac{E}{W}p_{c.m.} 
\label{pmin_pp}
\end{eqnarray}
\begin{eqnarray}
p_{max}(pp)= \frac{P}{W}E_{c.m.}+ \frac{E}{W}p_{c.m.} 
\label{pmax_pp}
\end{eqnarray}
where $E=E_2+E_1$, $\vec{P}=\vec{p_2}+\vec{p_1}$ and $W=\sqrt{E^2-\vec{P}^2}$ are respectively the energy, the momentum and the mass of the scattering $p p$ pair. The momentum and energy of the spectator proton, in the c.m. frame of the $p p$ pair are
\begin{eqnarray}
p_{c.m.}= \frac{\sqrt{W^2-4m^2}}{2}
\end{eqnarray}
\begin{eqnarray}
E_{c.m.}= \sqrt{p^2_{c.m.}+m^2} 
= \frac{W}{2}
\end{eqnarray} 

As in the previous section, the singular part of the integral picks the low momentum components of the deuteron wave function, relies on on-shell elementary matrix elements and is maximum when  $p_{min}(pp)=0$. The principal part vanishes under the rescattering peak, and contributes little to its tails: the situation is the same as in the $np$ rescattering sector of the $^2$H$(e,e^{\prime}p)n$ reaction (see {\it e.g.} Fig.~2 of~\cite{La05}).

The dependency upon $t=(k-p_{\pi})^2$ of the elementary photo-production amplitude $T_{\gamma n}$ is fixed by the external kinematics (see graph~III in Fig.~\ref{graph}) and not by the internal kinematics in the loop integral. Therefore it can be safely factorized out the integrals and evaluated assuming that the target nucleon is at rest in the deuteron, in which case $s= 2m\nu + m^2$. The integrals can be performed analytically, following the method outlined in~\cite{La78}. I have checked~\cite{La05} that this approximation is very close  (within 10~\%) to the full evaluation of the integrals, in the rescattering peak region. This saves computing time and both the singular and principal parts have been retained: the proton rescattering peak is therefore slightly wider than the pion rescattering peak in Fig.~\ref{pi1}. 

The proton-proton scattering amplitude is taken as:
\begin{eqnarray}
T_{pp} = (m_2m_1|\alpha + i\gamma (\vec{\sigma_1}+ 
\vec{\sigma_2})\cdot \vec{k}_{\perp}
\nonumber \\
+\;\mathrm{spin-spin} \;\mathrm{terms}\;|m^{\prime}_pm_p) 
\label{NN}
\end{eqnarray}
where $\vec{k}_{\perp}$ is the unit vector perpendicular to the scattering plane.

Above 500 MeV, the central part $\alpha$ dominates. It is almost entirely
absorptive, and takes the simple form
\begin{equation}
\alpha = -\frac{Wp_{cm}}{2m^2} \;(\epsilon + i)\;\sigma_{NN} \;\exp[\frac{\beta_N}{2}t_r]
\label{abs_amp} 
\end{equation}
Where $t_r=(p^{\prime}-p_1)^2$ is the four momentum transfer at the $pp$ scattering vertex. In the forward direction its imaginary part is related to the total cross section $\sigma_{NN}$, while the slope parameter $\beta_N$ is related to the angular distribution of NN scattering at forward angles. I use the same values as in~\cite{La05}. Note that the difference in the norm of eqs.~(\ref{abs_amp}) and~(\ref{scat_pin}) comes from the choice of the norm of the spinors, $\overline{u}u=1$.

\subsection{Coplanar kinematics}

Fig.~\ref{pi1} exhibits the salient predictions of the model. It corresponds to a coplanar kinematics, that can be achieved by detecting the pion and one of the proton with two well shielded magnetic spectrometers in Hall A or Hall C at JLab for instance. It shows the ratio of the full cross section to the quasi-free cross section, as function of the polar angle of the slow nucleon, $\theta_R=\theta_1$, when its momentum, $P_R=|\vec{p_1}|$, is kept constant at 200 MeV/c (lower curve) or 500 MeV/c (upper curve). The mass of the pair made of the pion and the fast (second) nucleon is kept constant at the value $W=\sqrt{(p_2+p_{\pi})^2}= 2.896$~GeV that corresponds to the absorption of a 4~GeV photon by a nucleon at rest. The four momentum transfer is also kept constant at the value $t=(k-p_{\pi})^2= -3$~GeV$^2$ which corresponds to the emission of the pion around $90^{\circ}$ in the $\pi p_2$ c.m. frame.

At high recoil momentum, rescattering mechanisms dominate over the quasi-free contribution. The top of the peaks corresponds to kinematics where an on-shell pion or nucleon can be produced on a nucleon at rest ($p_{\rm min}=0$), in the rescattering amplitude. The width of the peaks reflects the Fermi motion of the target nucleon. The physical picture is the following. The pion (resp. proton) is photo-produced on a neutron at rest in deuterium, propagates on-shell and rescatters on the spectator proton, also at rest in deuterium.  Two body kinematics requires that the angle between the scattered pion (proton) and the recoiling proton is constant (strictly $90^{\circ}$ for $pp$ elastic scattering). Since the recoiling nucleon momentum is fixed, the angles which it makes with the total momenta $\vec{p_1}+\vec{p_{\pi}}$ or $\vec{p_1}+\vec{p_2}$ are also fixed: typically $70^{\circ}$. So, the $\pi p$ or the $pp$ rescattering peaks form a cone centered along the direction of the total momentum of the corresponding scattering pair. In coplanar kinematics, two peaks appear for each rescattering, depending whether or not the pion and the recoiling proton are emitted on the same side of the photon.  

The difference in the height of each of these two peaks reflects the rapid variation with the photon energy of the elementary pion photo-production Regge cross sections: $s^{2\alpha(t)-2}$. Although the mass of the pair made of the pion and the fast proton has been kept constant in Fig.~\ref{pi1}, the incoming photon must also provide the energy of the slow recoiling nucleon: this depends on its direction of motion. For instance, at the top of the $\pi p$ peak that is located at the left in Fig.~\ref{pi1} the photon energy is $E_{\gamma}=3.432$~GeV, while it is $E_{\gamma}=5.436$~GeV at the top of the peak at the right. Since, in the rescattering amplitude, the photo-production occurs on a nucleon at rest  the corresponding masses are respectively 2.705 and 3.328~GeV. The situation is the same at the top of the two $pp$ rescattering peaks.

\begin{figure}[h]
\begin{center}
\epsfig{file=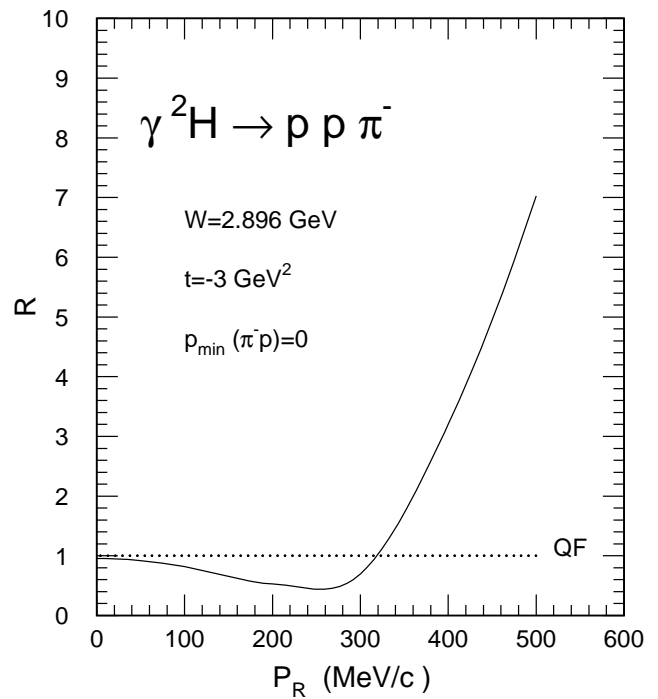,width=3.5in}
\caption{
Ratio of the total
to the quasi-free cross section  of the $^2$H$(\gamma,\pi^-p)p$ reaction
against the momentum of the recoiling proton, at the top of the
 $\pi p$ rescattering peak. The full line includes $\pi p$ scattering and $pp$ scattering (small effect). The dotted curve corresponds to the quasi-free process.}
\protect\label{pi2}
\end{center}
\end{figure}

At low recoil momentum, this effect is less dramatic since the energy difference is less important (it vanishes at $P_r= 0$!). Here  the rescattering amplitudes interfere destructively with the quasi-free amplitude, consistently with unitary. Since the elementary $\pi p$ and $pp$ scattering amplitudes are dominantly absorptive in the energy range covered by this study, a part of the strength is shifted from the quasi-free channel to inelastic channels. Above $p_r=300$~MeV/c, rescattering contributions take over and dominate the cross section. Figs.~\ref{pi2}  shows  this evolution of  the cross section
at the top of the $\pi p$ rescattering peak ($\theta_R=-50^{\circ}$ in Fig.~\ref{pi1}) with the recoil  momentum
$P_R$. 

\subsection{CLAS kinematics \label{clas}}

The CLAS~\cite{CLAS} set up at JLab allows to record events in the full available phase space and is well suited  to make a survey of the cross section of the $^2$H$(\gamma,p\pi^-)p$ reaction and to exploit its features which we just discussed. 

Three superconducting coils generate a toroidal field perpendicular to the photon beam axis and define six sectors where particles are detected by wire chambers and scintillators. The geometrical fiducial acceptance represents more than $2\pi$~sr. It covers a range of polar angles  between $11^{\circ}$ and $140^{\circ}$, but the coils define six azimuthal regions where the detector is blind.

I have implemented the code which computes the cross section of the $^2$H$(\gamma,p\pi^-)p$ reaction in a Monte-Carlo code which generates events in the full fiducial acceptance of CLAS. I sample, with a flat distribution, the three-momentum $\vec{p_1}$ of the slow proton and the two angles $\cos\theta_2$ and $\phi_2$ of the fast proton. If each proton falls in the fiducial acceptance, which I take from ref.~\cite{Nic04}, I record the kinematics of the event in a database (namely an Ntuple in the CERN package PAW~\cite{PAW}) and I weight it with the corresponding differential cross section
\begin{eqnarray}
\frac{d\sigma}{dp_1d\Omega_{p_1}d\Omega_{p_2}}=J\times
\frac{d\sigma}{d\vec{p_1}[d\Omega_{\pi}]_{cm2}}
\end{eqnarray}
where $J$ is the relevant Jacobian
\begin{eqnarray}
J= \frac{Q_f  |\vec{p_2}|^3|\vec{p_1}|^2}
{\mu_{c.m.}|E_{\pi}|\vec{p_2}|^2 -E_2\vec{p_{\pi}}\cdot\vec{p_2}|}
\end{eqnarray}
(see section~\ref{model} for the definition of momenta and energies.)

The events in the data base are then binned as the experimental data, with the same cuts. This is the most straightforward way to compare a theory with experiments, or to simulate experiments, that are carried out over a wide and complicated phase space. 

\begin{figure}[h]
\begin{center}
\epsfig{file=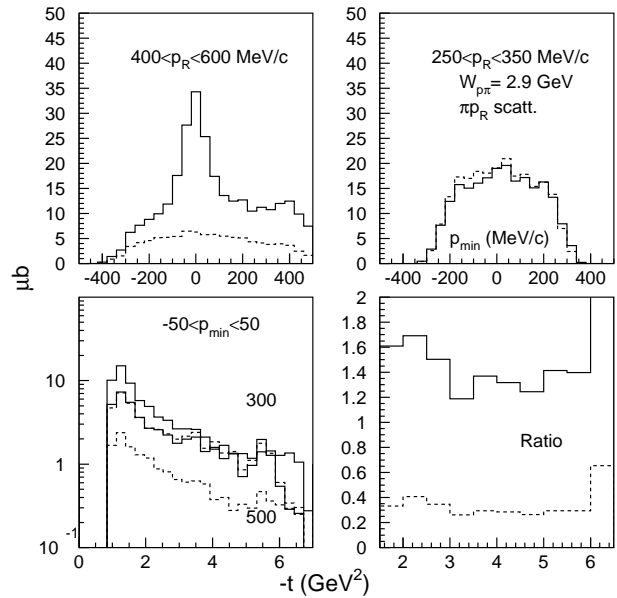,width=3.5in}
\caption{CLAS kinematics for $\pi p$ rescattering in the $^2$H$(\gamma,pp_R)\pi^-$ reaction. The beam end point is 6 GeV. The full histograms correspond to the full calculation, while the dashed histograms correspond to the quasi-free process only. See text for the description of the cuts which have been used in each window.}
\protect\label{panel_ppi}
\end{center}
\end{figure}

Figs.~\ref{panel_ppi}  shows various observables which emphasize the pion nucleon rescattering sector. The real photon beam end point has been set to $E_{\gamma}=6$~GeV, but the mass of the fast proton pion pair has been restricted to the range $2.85<W_{p_2\pi}<2.95$~GeV to keep contact with Fig.~\ref{pi1}. Only events corresponding to large momentum transfer $-t>1$~GeV$^2$ has been retained. The panel shows cross sections integrated over the various bins, within the CLAS fiducial acceptance. The top parts show the distribution of the minimum momentum $p_{min}(p_1\pi)$, eq.~\ref{pmin_ppi}, of the spectator proton, in the pion nucleon scattering loop, for which the pion can propagate on-shell. On the left, the cut $400<P_R<600$~MeV/c has been applied on the momentum $p_1$ of the slow nucleon: the pion nucleon rescattering peak clearly appears at $p_{min}=0$. On the right, the cut $250<p_R<350$~MeV/c has been applied: rescattering effects are small here, consistently with Fig.~\ref{pi2}, and the shape of the distribution reflects the kinematics and the detector acceptance. This is a good reference point which emphasizes the quasi-free process.

A further cut $-50<p_{min}(p_1\pi)<50$~MeV/c has been applied in the bottom parts of Fig.~\ref{panel_ppi}: it emphasizes pion nucleon rescattering. The $t$ distribution is plotted on the left for either low recoil momentum ($\sim 300$~MeV/c) or the high recoil momentum ($\sim 500$~MeV/c) bands. The ratio of second ($p_R=500$~MeV/c) to the first ($p_R=300$~MeV/c) $t$ distribution is plotted on the right. In plane wave, it is nothing but the ratio of these high momentum to low momentum  components of the nucleon momentum distribution in deuterium. The full ratio is really a measure of the evolution of the top of the pion nucleon rescattering peak with the four momentum transfer $t$, which fixes the hard scale. It is almost flat (the oscillations are due to the statistical accuracy of the Monte Carlo sampling) and provides us with a good starting point to look for deviations, especially at high $-t$, which could reveal the onset of color transparency for instance. The last bins ($-t>6$~GeV$^2$) in the $t$ distribution  should be disregarded since they correspond to the kinematical limits where the detector acceptance differs strongly at low and high recoil momentum.

\begin{figure}[h]
\begin{center}
\epsfig{file=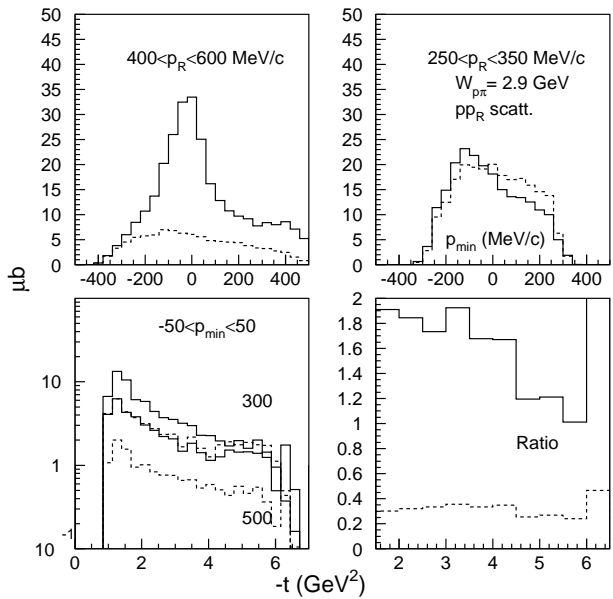,width=3.5in}
\caption{CLAS kinematics for $pp$ rescattering in the $^2$H$(\gamma,pp_R)\pi^-$ reaction. The beam end point is 6 GeV. The full histograms correspond to the full calculation, while the dashed histograms correspond to the quasi-free process only. See text for the description of the cuts which have been used in each window.}
\protect\label{panel_pp}
\end{center}
\end{figure}

Fig.~\ref{panel_pp} shows the same observables in the proton proton rescattering sector. Now, the minimum momentum $p_{min}(p_2p_1)$ is the lowest value, eq.~\ref{pmin_pp}, of the momentum of the spectator proton for which the other proton can propagate on-shell in the nucleon nucleon scattering loop. Again, the bins at the highest values of $-t$ should be disregarded since they lie at the kinematics limits. Also the statistical accuracy can be improved by running the Monte Carlo code with more events (but also longer!).

\begin{figure}[h]
\begin{center}
\epsfig{file=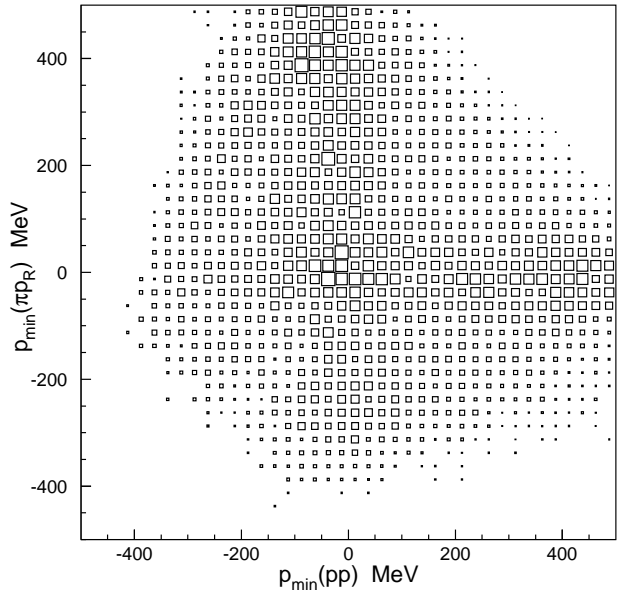,width=3.5in}
\caption{The joint distribution of singularities in $pp$ and $p\pi$ rescattering in the $^2$H$(\gamma,pp_R)\pi^-$ reaction. The beam end point is 6 GeV. The range of mass of the fast $p \pi$ pair is $2.85<W_{p_2\pi}<2.95$~GeV. The range of the four momentum transfer is $-t>1$~GeV$^2$. The range of the momentum of the slow proton is $400<P_R<600$~MeV/c.}
\protect\label{sing_bidim}
\end{center}
\end{figure}

In Fig.~\ref{panel_ppi}, $pp$ rescattering gives also a contribution below the $p\pi$ rescattering peak. Also, $p\pi$ rescattering gives a contribution under the $pp$ rescattering peak in Fig.~\ref{panel_pp}. These contaminations can be removed by cutting the overlapping region in the joint distribution of the rescattering singularities which is shown in Fig.~\ref{sing_bidim}.

In the Monte Carlo simulation, the choice have been made to detect the two protons. The advantage is that the efficiency to detect a proton in each sector of CLAS is very good (better than $90$~\%, and only a small correction has to be applied to the histograms before comparing them to experiment): this is particularly interesting when one selects events corresponding to large recoil momentum, of which the probability is small. But this prevents to record events with small recoil momentum, since   CLAS cannot detect protons with momentum lower than $\sim 250$~MeV/c. For recording such events one must detect the $\pi^-$, which is bent inward, in the beam direction, by the magnetic field and also decay in flight. Its detection efficiency is much smaller that  the one of a proton, but on the other hand the cross section is higher at low recoil momentum.

\subsection{Determination of the $\pi^-$ elementary production amplitude}

\begin{figure}[h]
\begin{center}
\epsfig{file=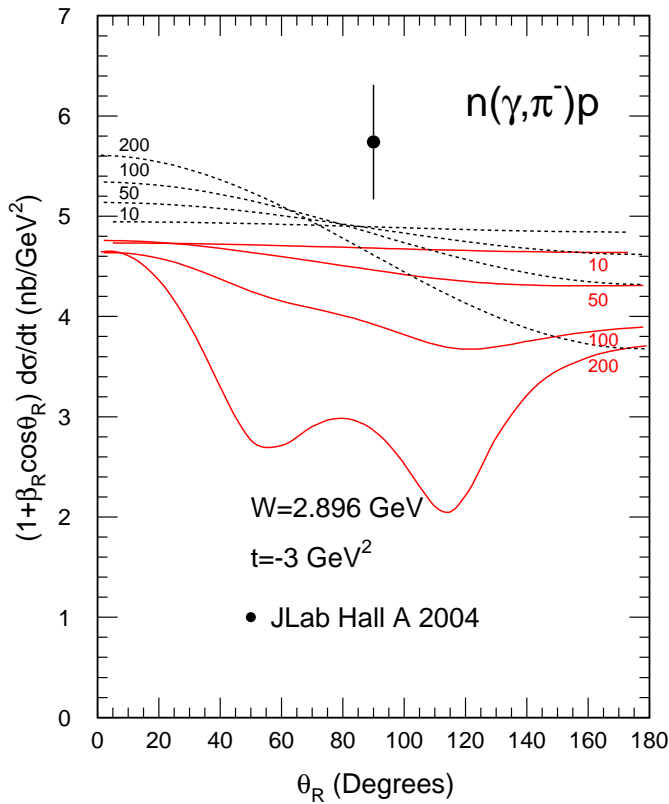,width=3.5in}
\caption{Color on line. The c.m. cross section  of the $n(\gamma,\pi^-)p$ reaction as extracted  from the analysis of the $^2$H$(\gamma,\pi^-p)p$ reaction when the angle of the recoiling proton varies but its momentum is kept
constant at 10, 50, 100 and 200 MeV/c. The dotted curves correspond to the quasi-free process. The kinematics is coplanar, and only the part of the angular distribution, which corresponds to the emission of the pion and the recoiling proton on different side of the photon, is shown. Each curve is labeled by the value of the corresponding recoil momentum; on the left for the quasi-free result, on the right for the full calculation}
\protect\label{low_p}
\end{center}
\end{figure}

All the preceding discussion relies on the good knowledge of the cross section of the elementary reaction $n(\gamma,\pi^-)p$. It can be determined in the same data set~\cite{g10}, since no free neutron target exists. To that end, one has to detect the $\pi^-$ and the fast proton $p_2$, and restrict the analysis to small values of the spectator slow proton, let say $p_1< 100$~MeV/c.

Fig.~\ref{low_p} shows that rescattering corrections are less than $10$~\% up to recoil momentum around $50$~Mev/c. Above, the effects are larger. The neutron cross section  is deduced from the deuterium cross section as follows
\begin{eqnarray} 
(1+\beta_1 \cos\theta_1)\frac{d\sigma}{d\Omega_{\pi}}= 
\frac{d\sigma}{d\vec{p_1}d\Omega_{\pi}}
\times \frac{1}{\rho(|\vec{p_1}|)}
\label{elem_Xsec}
\end{eqnarray}
where $\rho(|\vec{p_1}|)$ is the neutron momentum distribution in the deuteron. I kept the flux factor in order to quantify the corresponding slope of the quasi-free cross section.

The kinematics, $W=2.896$~MeV and $t=-3$~GeV$^2$, corresponds to one of the few data point~\cite{Zh04} recently measured in Hall A at JLab. It corresponds to the kinematics in Fig.~\ref{pi1}, as well as the average kinematics covered by this study, and indicates that the elementary amplitude is not far off the mark.
This experimental datum (and few others) has been obtained from the analysis of the $^2$H$(\gamma,\pi^-)$ reaction induced by a Bremsstrahlung photon beam, which averages the elementary cross section over the Fermi motion of the neutron in the deuterium target. Since the major contribution comes from neutron momentum in the range of $\sim 80$~Mev/c, interaction effects  cannot be neglected. The analysis of the high statistics Hall B experiment~\cite{g10}, along the line of Fig.~\ref{low_p}, would be very welcome in order to enlarge the data set and improve its accuracy.

In order to keep contact with Fig.~\ref{pi1}, the cross sections at $p_R=p_1= 200$~MeV/c are also plotted in Fig.~\ref{low_p}. The ratio of the full calculation to the quasi-free one is plotted in Fig.~\ref{pi1}. When multiplied by the momentum distribution $\rho(|\vec{p_1}|)$ these curves become the fully differential cross sections $d\sigma/d\vec{p_1}d\Omega_{\pi}$. One notes that the quasi-free cross sections follow almost exactly the variation of the flux factor $1+\beta_1 \cos\theta_1$, but decreases a little when the recoil momentum $p_1$ increases, by 2\% at 100~MeV/c, 5\% at 200~MeV/c and no more than 10\% above. This is a consequence of the choice of the off-shell extrapolation of the elementary pion photo production amplitude (see Appendix) and a measure of the uncertainty on the determination of the quasi-free amplitude at high recoil momenta. However, rescattering amplitudes dominate here (by about a factor 5 and more, at the top of the rescattering peak) and they are driven by on-shell elementary matrix elements and the low momentum components of the nuclear wave function. Therefore the uncertainties in the off-shell extrapolation of the elementary amplitudes  do not affect the cross section in the domains of the rescattering peaks. 

\section{The strange sector \label{strange}}

The extension to the $^2$H$(\gamma,K^+\Lambda)n$ reaction is straightforward. The amplitude of the elementary reaction $p(\gamma,K^+)\Lambda$ is driven by the exchange of the Regge trajectories of the $K$ and the $K^*$~\cite{La97}. Besides trivial changes in the mass of the particles, the Regge trajectories and the coupling constants, the reaction amplitudes exhibit the same form as in the pion production sector (section~\ref{pi}).

The kaon is produced on the proton. Since the neutron and the $\Lambda$ in the final state are distinct particles, there is no need to antisymmetrize the reaction amplitude. 

All the coupling constants and Regge propagators are given in ref.~\cite{La97}.  I use the slope parameters $\beta_K=3$~GeV$^{-2}$~\cite{La72} and $\beta_\Lambda=\beta_{pn}$~\cite{La05}, and the cross sections~\cite{PDG}: $\sigma_{K^+n}= 18$~mb and   $\sigma_{\Lambda n}=35$~mb in the $K^+n$ and $\Lambda n$ scattering amplitudes respectively. However these parameters are less known than in the $\pi N$ and $NN$ scattering sector, and their choice should be refined by the analysis of the $^2$H$(\gamma,K^+\Lambda)n$ reaction at low four momentum transfer $t$.

\begin{figure}[h]
\begin{center}
\epsfig{file=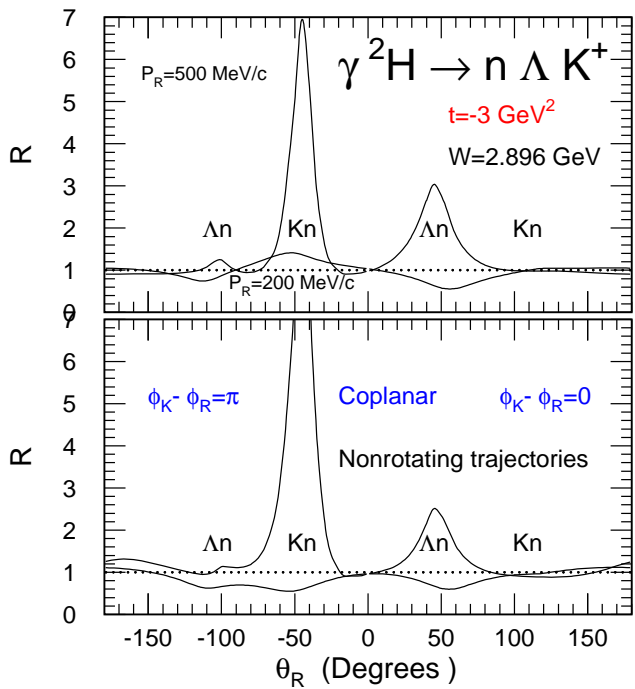,width=3.5in}
\caption{Color on line. Ratio of the total
to the quasi-free cross section  of the $^2$H$(\gamma,K^+\Lambda)n$ reaction
against the angle of the recoiling neutron whose the momentum is kept
constant at 500 MeV/c (top curves) and 200 MeV/c (bottom curves). The peaks labeled $K^+n$ and $\Lambda n$  correspond respectively  to $K^+n$ rescattering and $\Lambda n$ on shell rescattering. The dotted curves correspond to the quasi-free process. The kinematics is coplanar, and positive angles correspond to the emission of the kaon and the recoiling neutron on the same side of the photon. The top panel corresponds to a rotating phase, while the bottom panel corresponds to no phase in Regge trajectories. }
\protect\label{kaon}
\end{center}
\end{figure}

Fig.~\ref{kaon} shows the ratio of the full cross section to the quasi-free cross section, as function of the polar angle of the slow neutron, $\theta_R=\theta_n$, when its momentum, $P_R=|\vec{n}|$, is kept constant at 200 MeV/c (lower curve) or 500 MeV/c (upper curve). The mass of the pair made of the $K^+$ and the $\Lambda$ is kept constant at the value $W=\sqrt{(p_{\Lambda}+p_{K^+})^2}= 2.896$~GeV that corresponds to the absorption of a 4~GeV photon by a nucleon at rest. The four momentum transfer is also kept constant at the value $t=(k-p_{K^+})^2= -3$~GeV$^2$ which corresponds to the emission of the kaon around $90^{\circ}$ in the $K^+ \Lambda$ c.m. frame.

At high recoil momentum, the pattern is the same as in the pion production sector (Fig.~\ref{pi1}). The height of the  rescattering peaks are different simply because the hadronic cross sections and their slopes are different. At low recoil momentum however, the pattern is different in the $Kn$ scattering sector. The reason is that contrary to the $n(\gamma,\pi^-)p$ reaction Regge amplitudes, the $p(\gamma,K^+)\Lambda$ reaction Regge amplitudes exhibit a phase, $\exp [-\imath \pi \alpha(t)]$ (see the discussion in section~2.3.2 of ref.~\cite{La97}). In the $Kn$ rescattering integral, which selects nucleons almost at rest in deuterium, the average four momentum transfer $t$ is different than in the quasi-free production amplitude, where the target nucleon moves with a momentum fixed by the kinematics. This difference in $t$ changes the phase and compensates the destructive interference, between the quasi-free amplitude and the absorptive rescattering amplitude, around $200$~MeV/c. When the Regge phases are turned off, in the bottom part of Fig.~\ref{kaon}, the pattern becomes the same as in the $\pi^-$ production channel, in Fig.~\ref{pi1}.

This effect does not occur in the $\Lambda n$ rescattering sector, since $t$ is defined by the external kinematics and is the same in the rescattering amplitude and the rescattering amplitude (See Fig.~\ref{graph}).  

\begin{figure}[h]
\begin{center}
\epsfig{file=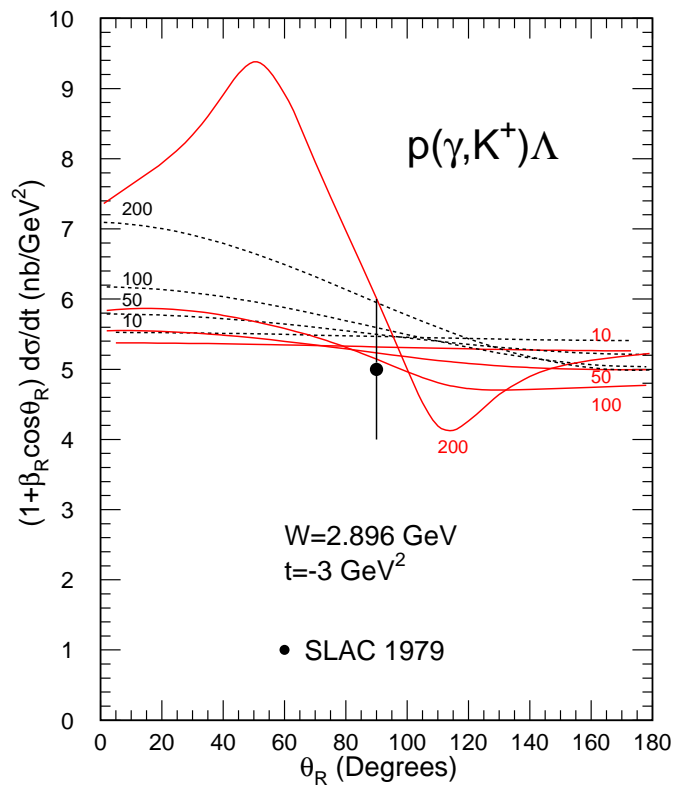,width=3.5in}
\caption{Color on line. The c.m. cross section  of the $p(\gamma,K^+)\Lambda$ reaction as extracted  from the analysis of the $^2$H$(\gamma,K^+\Lambda)n$ reaction when the angle of the recoiling neutron varies but its momentum is kept
constant at 10, 50, 100 and 200 MeV/c. The dotted curves correspond to the quasi-free process. The kinematics is coplanar, and only the part of the angular distribution, which corresponds to the emission of the Kaon and the recoiling neutron on different side of the photon, is shown. Each curve is labeled by the value of the corresponding recoil momentum; on the left for the quasi-free result, on the right for the full calculation}
\protect\label{low_p_kaon}
\end{center}
\end{figure}

Fig.~\ref{low_p_kaon} shows the value of the elementary kaon photo production cross section as extracted from a deuteron target with eq.~\ref{elem_Xsec} for different values of the recoiling neutron. As in the pion production sector the quasi-free cross section follows the variation of the flux factor $1+\beta_1 \cos\theta_1$ and departs from the free cross section ($p_R=10$~MeV/c) when the recoil momentum increases. The datum in Fig.~\ref{low_p_kaon} is one of the few experimental cross sections available at high momentum transfer~\cite{An76}

The reaction amplitudes rely on the elementary photo-production of kaon on a proton target. The Regge model leads to a fair agreement with the existing set of data around $E_{\gamma}= 4$~GeV (see~\cite{La97}). This data set can be greatly enlarged by the analysis of the high statistical accuracy experiment~\cite{g11} recently completed in  CLAS, with a beam of real photon of $4$~GeV on a proton target.

To achieve the full exclusivity of kaon production off a deuterium target one needs to detect both the $K^+$ and the $\Lambda$, which can be identified by its decay into a $\pi^-p$ pair. CLAS is ideally suited to record such a three charged particle configuration. When the $\Lambda$ decay distribution is implemented in the Monte Carlo code, the simulation leads to histograms similar to those which have been obtained in the pion sector. Since the physical content is the same, they are not shown but will be compared with experiment with the same cuts when the analysis is completed.

Other channels can also be studied. Of particular interest is the $^2$H$(\gamma,K^{\circ}\Lambda)p$ reaction~\cite{Pav05} where all the (decay) particles in the final state are charged. As in the $\pi^-$ sector the cross section of the elementary reaction $n(\gamma,K^{\circ})\Lambda$ must be determined from the same data set, demanding a low momentum ($p<50$~MeV/c) spectator proton. Also $^2$H$(\gamma,K^+\Lambda^*)n$ channel should be considered, since a Regge model based on the exchange of the $K$ and $K^*$ mesons~\cite{Gui96} leads also to a good agreement of the differential cross-section of the elementary reaction $p(\gamma,k^+)\Lambda^*$ at $E_{\gamma}=3.5$~GeV. Any signal in the $K^+n$ scattering sector should be the same as in the $^2$H$(\gamma,K^+\Lambda)n$ reaction.

\section{Vector meson production \label{vector}}

Exclusive Vector Mesons production on few body systems is certainly very
promising.  It allows  to prepare a pair  of quarks, with an  adjustable
transverse size,  and to  study its  interaction with  a nucleon in well
defined kinematics.  Furthermore,  the coherence time (during  which the
incoming  photon  oscillates  into  a  $q\overline{q}$  pair)  and   the
formation time (after  which this pair recombines into the  final meson)
can be  adjusted independently  to the  internucleonic distance.   

A  special  emphasis  should  be  put  on  $\phi$  and  $J/\Psi$  mesons
production:  not only these narrow states are more easy to identify, but
their flavor content,  different from that  of the ground  state of cold
hadronic matter, makes them a promising probe.

\begin{figure}[h]
\begin{center}
\epsfig{file=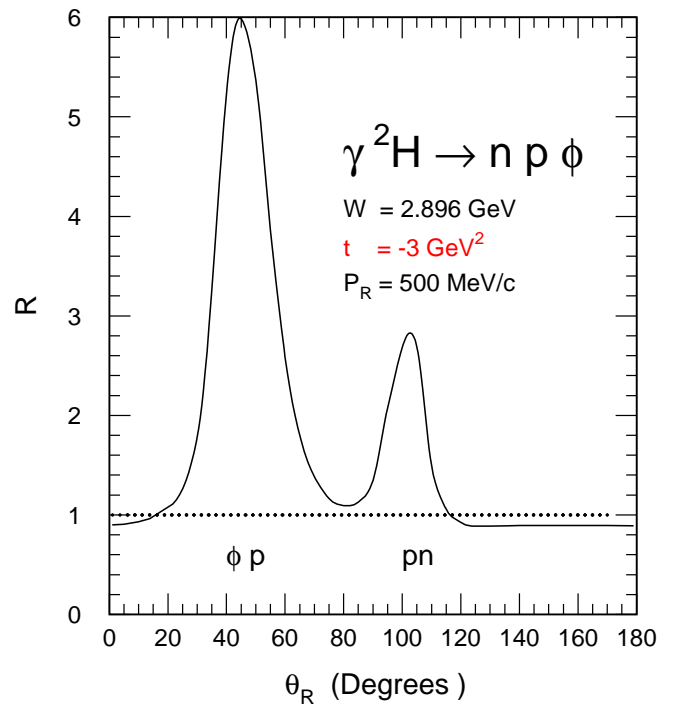,width=3.5in}
\caption{Color on line. Ratio of the total
to the quasi-free cross section  of the $^2$H$(\gamma,\phi p)n$ reaction
against the angle of the recoiling neutron whose the momentum is kept
constant at 500 MeV/c. The kinematic is coplanar and only the part, that corresponds to the emission of the meson and the recoiling neutron on different sides of the photon, is shown.}
\protect\label{phi}
\end{center}
\end{figure}

Fig.~\ref{phi} shows the expected signal in the $\phi$ photo-production channel, when the photon beam end point is $E_{\gamma}=6$~GeV. The model is a straightforward extension of the previous amplitudes. Again, the mass of the $p\phi$ fast pair and the four momentum transfer are set at $W_{p\phi}= 2.896$~MeV and $t=(k-p_{\phi})^2=-3$~GeV$^2$ respectively. The elementary amplitude~\cite{La00b,Ca02} is based on the exchange of two non perturbative gluons and uses a correlated nucleon wave function. It leads to a very good account of the $p(\gamma,\phi)p$ reaction~\cite{An00} recently measured at JLab at $E_{\gamma}=4$~GeV. The $\phi$ can be photo-produced on the proton as well as on the neutron: this has been taken into account in the model. The $pn$ scattering amplitude is defined according to ref.~\cite{La05}, while the $\phi n$ total cross section and slope parameter are respectively $\sigma_{\phi n}=20$~mb and $\beta=6$~Gev$^{-2}$. Again those quantities are almost unknown, and must be determined by the analysis of the  $^2$H$(\gamma,\phi)pn$ reaction at low $-t$. The $\phi$ nucleon scattering cross section has been extracted from one experiment (see ref.~\cite{VMD}), while I have taken the universal slope for high energy diffractive processes.

Such  a study may end up with a better understanding of the formation of  vector
mesons in  cold hadronic  matter, and will be a reference for the study of  vector meson  production in  heavy ion  collisions.

\section{Color Transparency and hadrons propagation \label{ct}}

The expected  effect of Color  Transparency would be a reduction of the rescattering  peak in Figs.~\ref{panel_ppi}, \ref{panel_pp}, \ref{kaon} and~\ref{phi} when the four momentum transfer $-t$ increases. The idea is that hard scattering mechanisms produce colorless dipoles with a small transverse size. Their scattering cross section is therefore expected to be reduced according to the square of the ratio of their transverse size to the transverse size of their asymptotic states. 

In addition, the rescattering peaks are expected to be wider since the small configuration is not an eigenstate of the mass operator. It is rather a combination of particles which can be diffractly excited from the ejectile~\cite{Ni94}. In the rescattering integrals (eqs.~\ref{pi_rescat} and \ref{p_rescat}) the propagator of these excited states should be added to propagator of the meson or the baryon which rescatters. The corresponding singularities are closely related to the mass of each particle of the spectrum and lead to peaks at slightly different locations. The rescattering peaks will be spread according to the actual distribution  of these states in the mass spectrum of the  small configuration. 

In previous reviews or prospect talks~\cite{La98,La98a,La00}  I have used a toy model based on a geometrical expansion of the  mini configuration of the  ejected hadron. Now, it is superseded by the quantum diffusion model~\cite{Fr94,Ja96,Ni94,Ni92} which can be implemented in eqs.~\ref{pi_rescat} and \ref{p_rescat}. I defer this study until dedicated experiments are performed.   Only experiments  will tell us what is the relevant nature  of the process which governs the formation and the evolution of such a small configuration.

This paper provides a base line calculation, under the assumption that normal particles rescatter. This is a unique situation, which relies on the evolution of a peak rather than on the level of attenuation of a flat cross section.

The key parameter is the time scale of the interaction. If it is long, the degrees of freedom are hadrons, and  well defined unitary peaks appear according to the study presented in this paper. If it is short enough, the degrees of freedom are quarks or exotic objects and  unitary peaks should be different or disappear.

Exclusive reactions at  hight $t$  are certainly best suited to these studies: they define the small interaction volume where quarks may be the relevant degrees of freedom. The virtuality $Q^2$ of the photon exchanged in electron scattering defines the volume of observation. Playing with this two independent hard scales is the key to the understanding of these rare processes~\cite{La04}. It can be started at 6 GeV, but clearly higher energies (12 GeV and even above) are needed

\section{Exotics \label{pent}}

The rescattering peaks also offer us with a tool to determine the cross section of the interaction with nucleon of  unstable or exotic particles, of which a beam is not available. Obvious examples are the photo-production of strange baryons ($\Lambda, \Sigma, \ldots$) of vector mesons ($\phi, J/\Psi, \ldots$).

They can also offer us with a way to disentangle elusive exotics, like pentaquarks for instance, and the physical background. One of the way to chase pentaquarks has been to determine the variation of the mass of the $K^+ n$ pair which is emitted in photo-reactions induced on nuclear~\cite{LEPS} or deuterium~\cite{Ste03} targets. Selecting the $K^+ n$ unitary rescattering peak, $|p_{min}(n K^+)|<100$~MeV/c for instance, in the reactions $^2$H$(\gamma,K^+\Lambda)n$ or $^2$H$(\gamma,K^+\Lambda^*)n$ would be the best way to master the physical background in the mass distribution of the  $K^+ n$ pair. 

The extension of the model presented in this paper is straightforward: simply the high energy description of the $K^+ n$ scattering should be replaced by a low energy description consistent with the existing data set. The contribution of resonant state, of a given width, could be added in order to set a limit of the production cross section of a possible exotic. Also, the decay distribution of the $\Lambda$ of the $\Lambda^*$ should be taken into account in the Monte Carlo simulation of the experiment. I defer this study to a forthcoming paper~\cite{La06}.

\section{Conclusion \label{conc}}

At the top of each unitary rescattering peak, the reaction mechanism is well under control. It depends on on-shell elementary matrix elements and involves the low momentum components of the nuclear wave function. This is a good starting point to access the interaction with nucleons of exotic objects and short lived particles.

This offers us a last chance to see and study Color Transparency in the present JLab energy range, more particularly in the strange quark sector. It also gives access to the determination of the cross sections of the scattering of vector and pseudo-scalar mesons on nucleons in a cold nuclear environment. Their knowledge is a key to the analysis of collisions between heavy ions at high energy.

For simplicity, all the numerical predictions in this study rely on a high energy description of the elementary matrix element. The photo-production amplitude is described by the exchange of the Regge trajectories of pseudo-scalar and vector mesons. The hadron scattering amplitudes are almost entirely absorptive. This treatment is already valid at kinematics that are achievable at JLab at $6$~GeV: mass of the meson nucleon system above $2$~GeV, relative kinetic energy between baryons above $0.5$ GeV.  It will be even more valid in the kinematical range that will be accessible when the CEBAF energy is upgraded to $12$~GeV.

The method can be easily adapted at lower energies, by implementing the relevant description (phase shift expansion, for instance) of the elementary amplitudes~\cite{La06}. It may prove to be useful to predict the physical background and the production cross section of elusive particles, such as pentaquarks for instance.

\section*{Acknowledgement}

I acknowledge the warm hospitality at JLab where this work was completed. The Southern Universities Research Association (SURA) operates Thomas Jefferson National Accelerator Facility (JLab) for the US Department of Energy under Contract No DE-AC05-84ER40150. Parts of this work have been done before I left Saclay. 

\section*{Appendix}

In ref.~\cite{La97}, the elementary photo-production amplitude has been expressed in terms of Dirac matrices and spinors. In this work, I have rewritten it in terms of Pauli matrices and spinors.

For on-shell nucleons, both expressions are equivalent. Both are Lorentz and gauge invariant. They are valid in any frame.

For off-shell nucleons, I made the choice to conserve the three momenta at each vertex, to conserve the energy in the invariant operator but to use the on-shell energy $E=\sqrt{p^2+m^2}$ in the normalization, $\sqrt{E+m}$, and the denominator, $E+m$, of Pauli spinors. This choice follows the time ordered expression of Feynman diagrams~\cite{Gro74}.

For off shell nucleons, gauge invariance is lost. However, in the rescattering peaks the target nucleon is almost at rest and the kicked nucleon is on shell. Therefore, the electromagnetic current is conserved in the dominant amplitudes that are considered in this paper.  

For the sake of completeness I reproduce the demonstration~\cite{La77} that I gave about thirty years ago. The quasi free matrix element takes the form:
\begin{eqnarray}
{\cal M}_I= - \overline{u}(p_2,m_2) \Theta
\frac{\gamma\cdot n +m}{n^2- m^2} \overline{u}(p_1,m_1)
\Gamma_{\mu}(p_D,n) \phi_D^{\mu}(M)
\nonumber \\
{}
\end{eqnarray}
where $\Theta$ is the elementary photo-production operator, $\Gamma_{\mu}$ the $^2$H$np$ vertex function and $\phi_D^{\mu}$ the deuteron field.

Retaining only the positive energy part of the neutron propagator~\cite{Gro74}  and neglecting its negative energy part, one gets:
\begin{eqnarray}
{\cal M}_I=  -\frac{m}{E_n}\sum_{m_n}
\frac{\overline{u}(p_2,m_2) \Theta u(\overline{n},m_n)
\overline{u}(\overline{n},m_n)\overline{u}(p_1,m_1)}{n^{\circ}-E_n}
\nonumber \\
\Gamma_{\mu}(p_D,n) \phi_D^{\mu}(M)
\nonumber \\ {}
\end{eqnarray}
with $E_n=\sqrt{\vec{n}^2+m^2}\neq n^{\circ}= E_D-E_1$ and $\overline{n}=(\overline{n}^{\circ}=E_n,\overline{\vec{n}}=\vec{n})$.

Identifying $\Gamma_{\mu}(p_D,n) \phi_D^{\mu}(M)m/E_n(n^{\circ}-E_n)$ with the deuteron wave function and defining 
\begin{eqnarray}
T_{\gamma n}(\vec{p_2},m_2,-\vec{p_1},m_n)=\frac{1}{-i}
\overline{u}(p_2,m_2) \Theta u(\overline{n},m_n)
\end{eqnarray}
one gets eq.~(\ref{q_f}). In terms of Pauli spinors, it takes the form:
\begin{eqnarray}
T_{\gamma n}(\vec{p_2},m_2,-\vec{p_1},m_n)=
(m_2|\imath\; \vec{\sigma}\cdot\vec{K}+ L|m_n)
\end{eqnarray}
which I use in this work. In the quasi-free amplitude, $\vec{n}=-\vec{p_1}$. In the rescattering amplitudes,  a similar expression takes into account the actual nucleon momenta.

The vector and scalar part of the $\pi$ exchange amplitude of the elementary reaction $n(\gamma,\pi^-)p$ are:
\begin{eqnarray}
\vec{K_{\pi}}= eg_{\pi NN}\sqrt{2}\frac{\sqrt{(E_n+m)(E_2+m)}}{2m} 
(t-m^2_{\pi})  {\cal P}^{{\pi}^-}_{\mathrm{Regge}}F_1(t)
\nonumber \\
\left[\left( \frac{\vec{n}}{E_n+m}-\frac{\vec{p_2}}{E_2+m}\right)
\left(\frac{(2\vec{p_2}-\vec{k})\cdot\vec{\epsilon}}{u-m^2}
-\frac{(2\vec{p_{\pi}}-\vec{k})\cdot\vec{\epsilon}}{t-m^2_{\pi}} \right)
\right.\nonumber \\
+\frac{\vec{\epsilon}}{u-m^2}\left(\nu-\frac{\vec{p_2}\cdot\vec{k}}{E_2+m}
-\frac{\vec{n}\cdot\vec{k}}{E_n+m}
\right. \nonumber \\ \left.
+\frac{\nu\vec{n}\cdot\vec{p_2}}{(E_n+m)(E_2+m)}\right)
\nonumber \\
+\frac{\vec{k}}{u-m^2}\left( \frac{\vec{n}\cdot\vec{\epsilon}}{E_n+m} 
+\frac{\vec{p_2}\cdot\vec{\epsilon}}{E_2+m} \right)
\nonumber \\
-\frac{\vec{n}}{u-m^2} \frac{\nu\vec{p_2}\cdot\vec{\epsilon}}{(E_n+m)(E_2+m)}
 \nonumber \\ \left.
-\frac{\vec{p_2}}{u-m^2} \frac{\nu\vec{n}\cdot\vec{\epsilon}}{(E_n+m)(E_2+m)}
 \right]\nonumber \\
{}
\end{eqnarray}
\begin{eqnarray}
L_{\pi}= eg_{\pi NN}\sqrt{2}\frac{\sqrt{(E_n+m)(E_2+m)}}{2m} 
(t-m^2_{\pi})  {\cal P}^{{\pi}^-}_{\mathrm{Regge}}F_1(t)
\nonumber \\
\frac{1}{u-m^2}
\left[ \frac{\vec{p_2}\times\vec{k}}{E_2+m} 
+\frac{\vec{k}\times\vec{n}}{E_n+m} 
-\frac{\nu\vec{p_2}\times\vec{n}}{(E_n+m)(E_2+m)}
\right]\cdot\vec{\epsilon}
\nonumber \\
{}
\end{eqnarray}
where $g^2_{\pi NN}/4\pi=14.5$, $t=(p_{\pi}-k)^2$ and $u=(p_2-k)^2$. I use the Regge propagator ${\cal P}^{{\pi^-}}_{\mathrm{Regge}}$ that corresponds to the $\pi$ saturating trajectory (as defined in ref.~\cite{La97}) and the dipole parameterization of the nucleon isovector form factor:
\begin{eqnarray}
F_1(t)=\frac{4m^2-2.79t}{(4m^2-t)(1-t/0.7)^2}
\end{eqnarray}

One recognizes the pure $\pi$ exchange amplitude (term in $1/(t-m^2_{\pi})$) and the part of the $u$-channel nucleon exchange amplitude (terms in $1/(u-m^2)$) which has been added to insure gauge invariance (see~\cite{La97}).

The vector and scalar part of the $\rho$ exchange amplitude of the elementary reaction $n(\gamma,\pi^-)p$ are:
\begin{eqnarray}
\vec{K_{\rho}}= e\frac{g_{\rho \pi\gamma}}{m_{\pi}}g_{\rho NN}\sqrt{2}
\frac{\sqrt{(E_n+m)(E_2+m)}}{2m}  {\cal P}^{\rho}_{\mathrm{Regge}}F_1(t)
\nonumber \\
\left\{\frac{\vec{p_2}\times\vec{n}}{(E_n+m)(E_2+m)}
\left[(1+\kappa_v)V^{\circ}
\right.\right.\nonumber \\
\left.
+\frac{\kappa_v}{2m}(V^{\circ}(n^{\circ}+E_2)-\vec{V}\cdot(\vec{n}+\vec{p_2}))
\right] 
\nonumber \\
\left.
(1+\kappa_v)\vec{V}\times\left[ \frac{\vec{p_2}}{E_2+m} 
-  \frac{\vec{n}}{E_n+m}\right]
\right\}
\nonumber \\
{}
\label{rho_vec}
\end{eqnarray}
\begin{eqnarray}
L_{\rho}= e\frac{g_{\rho \pi\gamma}}{m_{\pi}}g_{\rho NN}\sqrt{2}
\frac{\sqrt{(E_n+m)(E_2+m)}}{2m}  {\cal P}^{\rho}_{\mathrm{Regge}}F_1(t)
\nonumber \\
\left\{
(1+\kappa_v)\left[V^{\circ}\left(1+\frac{\vec{n}\cdot\vec{p_2}}{(E_n+m)(E_2+m)}
\right)
\right. \right.\nonumber  \\ 
\left.
-\vec{V}\cdot\left( \frac{\vec{n}}{E_n+m}+\frac{\vec{p_2}}{E_2+m}\right)
\right]
\nonumber \\
-\frac{\kappa_v}{2m} \left[1-\frac{\vec{n}\cdot\vec{p_2}}{(E_n+m)(E_2+m)} \right]
\nonumber \\
\left.
\left[ V^{\circ}(n^{\circ}+E_2)-\vec{V}\cdot(\vec{n}+\vec{p_2})\right]
\right\}
\nonumber \\
{}
\label{rho_scal}
\end{eqnarray}
I use the Regge propagator  ${\cal P}^{\rho}_{\mathrm{Regge}}$, with the saturating trajectory of the $\rho$ meson, and the coupling constants $g_{\rho \pi\gamma}=$~0.103, $g^2_{\rho NN}/4\pi=$~0.92 and $\kappa_v=$~6.1, as in ref.~\cite{La97}.

The four vector $V^{\mu}=(V^{\circ},\vec{V})$ contains the dependency upon the polarization vector $\vec{\epsilon}$ of the photon, in the following way:
\begin{equation}
\begin{array}{rrrrcr}
(-kR_y,& 0,& \nu R_z-kR^{\circ},& -\nu R_y)
&\mbox{if}& \vec{\epsilon}=(1,0,0)
\\
(kR_x,& kR^{\circ}-\nu R_z,&0,& \nu R_x)
&\mbox{if}& \vec{\epsilon}=(0,1,0)
\\
(0,& \nu R_y,& -\nu R_x,&0)
&\mbox{if}& \vec{\epsilon}=(0,0,1)
\end{array}
\end{equation}
where $R^{\circ}= \nu-E_{\pi}$ and $\vec{R}\equiv (R_x,R_y,R_z)= \vec{k}-\vec{p_{\pi}}$ are respectively the energy and the three momentum of the exchanged $\rho$ meson.

For the $p(\gamma,K^+)\Lambda$ reaction, the $K^*$ meson amplitudes take the same form as the $\rho$ meson exchange amplitudes~(\ref{rho_vec}) and~(\ref{rho_scal}), besides trivial changes in the masses, coupling constants and propagators.

The $K$ exchange amplitude contains the part of the $s$-channel nucleon exchange  amplitude which is strictly necessary to insure gauge invariance. It takes the form:
\begin{eqnarray}
\vec{K_{K}}= eg_{K N\Lambda}\sqrt{2}\frac{\sqrt{(E_n+m)(E_2+m_{\Lambda})}}
{\sqrt{4m m_{\Lambda}}} 
(t-m^2_{K})  {\cal P}^{K^+}_{\mathrm{Regge}}
\nonumber \\
\left[\left( \frac{\vec{n}}{E_n+m}-\frac{\vec{p_2}}{E_2+m_{\Lambda}}\right)
\left(\frac{(2\vec{n}+\vec{k})\cdot\vec{\epsilon}}{s-m^2}
+\frac{(2\vec{p_{K}}-\vec{k})\cdot\vec{\epsilon}}{t-m^2_{K}} \right)
\right.\nonumber \\
+\frac{\vec{\epsilon}}{s-m^2}\left(\nu-\frac{\vec{p_2}\cdot\vec{k}}{E_2+m_{\Lambda}}
-\frac{\vec{n}\cdot\vec{k}}{E_n+m}
\right. \nonumber \\ \left.
+\frac{\nu\vec{n}\cdot\vec{p_2}}{(E_n+m)(E_2+m_{\Lambda})}\right)
\nonumber \\
+\frac{\vec{k}}{s-m^2}\left( \frac{\vec{n}\cdot\vec{\epsilon}}{E_n+m} 
+\frac{\vec{p_2}\cdot\vec{\epsilon}}{E_2+m_{\Lambda}} \right)
\nonumber \\
-\frac{\vec{n}}{s-m^2} \frac{\nu\vec{p_2}\cdot\vec{\epsilon}}{(E_n+m)(E_2+m_{\Lambda})}
\nonumber \\ \left.
-\frac{\vec{p_2}}{s-m^2} \frac{\nu\vec{n}\cdot\vec{\epsilon}}{(E_n+m)(E_2+m_{\Lambda})}
 \right]\nonumber \\
{}
\end{eqnarray}
\begin{eqnarray}
L_{K}= eg_{K N\Lambda}\sqrt{2}\frac{\sqrt{(E_n+m)(E_2+m_{\Lambda})}}
{\sqrt{4mm_{\Lambda}}} 
(t-m^2_{K})  {\cal P}^{K^+}_{\mathrm{Regge}}
\nonumber \\
\frac{1}{s-m^2}
\left[ \frac{\vec{p_2}\times\vec{k}}{E_2+m_{\Lambda}} 
+\frac{\vec{k}\times\vec{n}}{E_n+m} 
-\frac{\nu\vec{p_2}\times\vec{n}}{(E_n+m)(E_2+m_{\Lambda})}
\right]\cdot\vec{\epsilon}
\nonumber \\
{}
\end{eqnarray}
where $n=(E_n,\vec{n})$ and $p_2=(E_2,\vec{p_2})$ now stand for the four momentum of the target proton and the outgoing $\Lambda$ respectively, and where $g_{K N\Lambda}^2/4\pi=$~10.6. 

As in ref.~\cite{La97}, I use the $K$ and $K^*$ linear trajectories in the Regge propagators, and no hadronic form factor ($F_1(t)=1$).

\end{document}